\documentclass[12pt,a4paper]{article}

\usepackage{geometry}
\usepackage{latexsym}
\usepackage{epsf}
\usepackage{amssymb}
\usepackage{graphicx}
\usepackage{amsmath, cite}
\usepackage{amsmath,amssymb,amsthm}
\usepackage{verbatim}
\usepackage{hyperref}
\renewcommand{\d}{\textrm{d}}

\newcommand{\w}{\wedge}

\newcommand{\SU}{\mathop{\rm SU}}

\newcommand{\kper}{k_{\perp}}
\newcommand{\kpar}{k_{\parallel}}

\renewcommand{\Im}{\operatorname{Im}}

\newcommand{\be}{\begin{equation}}
\newcommand{\ee}{\end{equation}}
\newcommand{\beq}{\begin{equation}}
\newcommand{\eeq}{\end{equation}}
\newcommand{\ba}{\begin{eqnarray}}
\newcommand{\ea}{\end{eqnarray}}
\newcommand{\bea}{\begin{eqnarray}}
\newcommand{\eea}{\end{eqnarray}}
\newcommand{\nn}{\nonumber}

\renewcommand{\d}{\textrm{d}}

\begin{document}
\numberwithin{equation}{section}

\begin{center}

{\LARGE \bf{$\mathcal{N}=1$ SUSY AdS$_4$ vacua in IIB SUGRA on group manifolds}}

\vspace{1.1 cm} { G. Solard}\\

\vspace{0.8 cm}{Laboratoire de Physique Th\'eorique et Hautes Energies,\\
Universit\'e Pierre et Marie Curie,\\ 4 Place Jussieu, 75252 Paris Cedex 05, France}\\

\vspace{0.8cm}

email : solard@lpthe.jussieu.fr

\vspace{1.2cm}

{\bf Abstract}
\end{center}

\begin{quotation}
We study $\mathcal{N}=1$ compactification of IIB supergravity to AdS$_4$. The internal manifold must have $\SU(2)$-structure. By putting some restrictions on the $\SU(2)$ torsion classes, we can perform an exhaustive scan of all possible solutions on group manifolds. We show that sourceless solutions don't exist and that the presence of O5 and O7 orientifolds is mandatory. We also find a new solution and relate it by T-duality to a known type IIA solution with O6 planes.
\end{quotation}
\newpage

\tableofcontents

\section{Introduction}

Even if  of no direct phenomenological interest, compactifications to four-dimensional Anti de Sitter space are worth studying for several
reasons.  For instance, they are relevant for the CFT$_3$/AdS$_4$ correspondence and they might represent the first step in the construction
of de Sitter vacua in string theory.
%Phenomenology seems to point to a positively curved universe. One way to obtain such a universe is to look for dS vacua. So one looks to obtain flux vacua breaking supersymmetry in order to obtain small positive cosmological constant. Unfortunately no such solution has been obtained in a clear and fully explicit way. In order to simplify the problem, one can break it in several part. This is why in this paper we try to tackle the problem of first obtaining SUSY AdS vacua. In this paper, we propose a systematic method to explore all group manifold and see if there exist a type IIB SUSY AdS vacuum on it. 
In type IIA the literature on  SUSY AdS$_4$  flux vacua is plentiful:  examples have been found both with  \cite{Derendinger:2004jn,Behrndt:2004km,DeWolfe:2005uu,Lust:2004ig,Caviezel:2008ik} or  without sources  \cite{Tomasiello:2007eq,Koerber:2010rn}.
Among the vacua with sources some  (see for instance  \cite{Dasgupta:1999ss,Giddings:2001yu} and their T-duals \cite{Grana:2006kf,Blaback:2010sj,Blaback:2012mu}) contain fully localized sources.  However  most  examples involve intersecting sources, D-branes or O-planes, which are  smeared in the transverse directions.  This raises the question of what is the meaning of 
a smeared orientifold plane and how such solutions can lift to full string theory  \cite{McOrist:2012yc,Douglas:2010rt}\footnote{Some partial result about O6 planes can be found in \cite{Saracco:2012wc}.}

The aim of this paper is to study  AdS$_4$ vacua in type IIB theory. These are less studied than their IIA counterparts. 
Some work in this direction can be found in \cite{Lust:2009zb,Caviezel:2009tu} and more recently in \cite{Petrini:2013ika}. 
A first question we aim at answering is  the existence of  AdS$_4$ SUSY vacua without sources. These clearly avoid all the aforementioned problems
about the possible validity of the solutions and are clearly important in the context of AdS/CFT. 
While supersymmetric AdS$_4$ vacua without sources are known in type IIA, there is only one known example in type IIB \cite{Lust:2009mb}.
Using $\SU(2)$ structure techniques\footnote{As shown in \cite{Lust:2004ig,Petrini:2013ika} $\mathcal{N}=1$ SUSY vacua in type IIB supergravity only exist on manifolds with SU(2) structure.},  the authors of \cite{Lust:2009zb} showed that, for constant warp factor and  a specific choice of $\SU(2)$ torsions, only non supersymmetric
sourceless vacua can be found.  One goal of this paper is to extend the analysis to a larger class of manifolds.  The most
general form of  the supersymmetry equations is too complicated to give general results. For this reason we will   focus  on group manifolds 
admitting an SU(2) structure and look for solutions with constant warp factor.  From simple manipulations of the traced Einstein equations it is easy to
see that in absence of sources the internal manifold must have positive curvature \cite{Petrini:2013ika}. This condition already rules out all nilmanifolds
as candidates for sourceless vacua in type IIB.  In this paper we will prove a stronger result:  under some restrictions on the SU(2) torsions (namely the vectors in the torsion classes are set to zero), it is not possible to have sourceless solutions. 
We will leave the analysis of the warped case for future work.

The same approach used to look for sourceless solutions, allows to scan all possible AdS$_4$  $\mathcal{N}=1$ SUSY vacua in type IIB on group manifolds. As before we will consider  constant warp factor and no torsions in the vector representation of $\SU(2)$. 
We use the pure spinors formalism developed in \cite{Grana:2006kf,Grana:2004bg,Grana:2005sn}. This permits to solve the SUSY equations using algebraic equations on the components of the fields on the internal manifold. In this way  we have been able to put constraints on the 
possible group manifolds and obtain an exhaustive list of  which manifolds admit  an AdS$_4$ vacuum. An important side result of this analysis is the fact that O5 and O7 planes are necessary in order to have such vacua. Unfortunately, the presence of intersecting orientifolds forces us to use smeared sources.

\vspace{0.3cm}

This paper is organized as follows. In Section \ref{susysect} we give the general conditions that the fields on the internal manifold have to satisfy in order to have $\mathcal{N}=1$ AdS$_4$. We also give the general form of $\SU(2)$-structure and rewrite the SUSY equations in this case. In Section \ref{looking}, we specialize to group manifolds and solve the SUSY equations in this case. We also discuss the charges coming from the RR Bianchi identities and we prove that no sourceless solution exist. In Section \ref{scanning}, we give the exhaustive list of group manifolds admitting an AdS$_4$ solution in our framework and we discuss a new solution on a solvmanifold. There we also prove that O5 and 
O7 orientifolds are necessarily present. 

Appendix \ref{structureconstants} gives the structure constants of the group manifold after having solved the SUSY equations. Appendix \ref{NSBI} gives the solutions to the NS Bianchi identity. Finally, in Appendix \ref{Tduality} we discuss how our new solution is related 
by T-duality to the IIA solutions given in \cite{Grana:2006kf}.

\section{The supersymmetry conditions}
\label{susysect}

We are interested in $\mathcal{N}=1$ SUSY AdS$_4$ vacua in type IIB theories. As discussed in \cite{Behrndt:2005bv, Lust:2009zb} such solutions are only possible
when the compactification manifolds have SU(2) structure group. Let us recall that a manifold  is said to be of SU(2) structure if it admits 
a complex one form $z$,  a real and a holomorphic two-form, $j$ and $\omega$, that are globally defined and satisfy
\bea
&& z \llcorner \bar{z} = 2\,, \qquad   z \llcorner z =  \bar{z} \llcorner \bar{z} = 0\,, \\
&& j \w \omega =0 \,, \\
&&  z \llcorner j  = z \llcorner \omega = 0 \,, \\
&& j \w j = \frac{1}{2} \omega \w \bar{\omega} \,.
\eea
Such forms can be constructed as bilinears of two globally defined spinors, $\eta_+$ and  $\chi_+ = \frac{1}{2} z \eta_-$ 
\bea
\label{SU2strf}
&& z_m =  - \chi^\dagger_- \gamma_m \eta_+ \,, \\
&& j_{mn} =  -i \eta^\dagger_ + \gamma_{mn} \eta_+ + i  \chi^\dagger_ + \gamma_{mn} \chi_+ \,, \\
&& \omega_{mn} = - i \chi^\dagger_+ \gamma_{mn} \eta_+ \,  .
\eea

\vspace{0.3cm}

In order to study $\mathcal{N}=1$ vacua with non trivial fluxes, it is convenient to use the language of  Generalized Complex Geometry \cite{Hitchin:2004ut,Gualtieri:2003dx}.
The idea is to express the ten-dimensional supersymmetry variations as differential equations on a pair of polyforms defined on the internal manifold.
In type IIB, one can decompose the ten-dimensional supersymmetry parameters as
\beq
\label{10dspinors}
\epsilon_i = \zeta_+ \otimes \eta^i_+ + \zeta_- \otimes \eta^i_-  \qquad \qquad i=1,2 \, ,
\eeq
where $\zeta_+$ and $\eta^i_+$ are Weyl spinors in four and six dimensions respectively\footnote{We choose the gamma matrices in such a way that
$\eta^i_- = (\eta^i_+)^\ast$ and $\zeta_- = (\zeta_+)^\ast$} and 
\bea
&& \eta^1_+ = a \eta_+ \,,\nonumber\\
&& \eta^2_+ = b (\kpar \eta_+ + \frac{1}{2} \kper  z_m \gamma^m \eta_-) \,,\label{6dspinors}
\eea
where $z =  z_m \gamma^m$ is the SU(2) structure one-form and $\eta_\pm$ are as in \eqref{SU2strf}. 
The complex functions $a$ and $b$ give the norms of $\eta^1$ and $\eta^2$
\beq
 \| \eta^1_+ \|^2 = |a|^2\,, \qquad \qquad  \| \eta^2_+\|^2= |b|^2 \, , 
 \eeq
  while the parameters $\kpar$ and $\kper$ ($\kpar^2 + \kper^2 = 1$) 
are related to the choice  of structure on the internal manifold.  
When  $\kpar = 0$ and $\kper=1$ the structure is $\SU(2)$, while the general case where both $\kpar$ and $\kper$ are non-zero is often referred to as dynamical $\SU(2)$  structure\footnote{When $\kpar = 1$ and $\kper=0$ the internal manifold is said to be of $\SU(3)$ structure. We will not consider
this case here.}.  When $\kpar$ and $\kper$ are non zero and constant, we speak of intermediate SU(2) structure rather than dynamical SU(2) structure \cite{Andriot:2008va}.

Then, one can use  the spinors $\eta_i$ to define two polyforms on the internal manifold $M$
\beq
\label{puresp}
\Phi_\pm  = \eta^1_+ \otimes \eta^{2 \dagger}_\pm\, ,
\eeq
whose explicit form depend on the choice of structure on $M$. In the most general case of intermediate 
SU(2) structure they are 
\bea
\label{Phim}
&& \Phi_-= -\frac{ab}{8}   z \wedge (\kper e^{ -ij }  + i \kpar \omega ) \, , \\
\label{Phip}
&& \Phi_+=  \frac{ a \bar{b}}{8}   \, e^{ z \bar{z} /2} (\kpar e^{-ij}  -i\kper \omega ) \, ,
\eea
where $z$, $j$ and $\omega$ are the forms defining the $\SU(2)$ structure.   The norm of the pure spinors $\Phi_\pm$ is related to the norm of
the  spinors $\eta^i$ by 
\beq
\label{normps}
\langle \Phi_\pm , \bar{\Phi}_\pm \rangle = - i \| \Phi_\pm \|^2 {\rm vol}_6 =  - \frac{i}{8}  | a |^2  |b|^2 {\rm vol}_6 \, ,
\eeq
where ${\rm vol}_6$ is the volume of the internal manifold and the product 
\beq
\label{Miukai}
\langle A , B  \rangle  = (A \wedge \lambda(B))|_{\rm top}
\eeq
is the Mukai pairing among forms, where $\lambda$ acts on a form as  the transposition of all indices
 \beq
 \label{lambdaop}
 \lambda(\omega_p) = (-)^{\lfloor p/2\rfloor} \omega_p \, .
 \eeq

\vspace{0.3cm}

As shown in \cite{Grana:2004bg},  for type IIB compactifications to AdS$_4$  the ten-dimensional supersymmetry variations  are equivalent to the following set of equations on the pure spinors $\Phi_\pm$ 
\bea
\label{eqm}
&&(\d - H  \w )(e^{2A - \phi} \Phi_-)  = - 2 \mu e^{A-\phi} {\rm Re } \Phi_+,  \\
\label{eqp}
&&  (\d - H \w )(e^{A -\phi } {\rm Re} \, \Phi_+) =0 \, , \\
\label{eqpb}
&& (\d - H \w )(e^{3 A -\phi } {\rm Im} \, \Phi_+) = -3e^{A-\phi}{\rm Im} \, (\bar{\mu}\Phi_-) -\frac{1}{8} e^{4 A} \ast \lambda(F) \,, 
\eea
 where $\phi$ is the dilaton, $A$ the warp factor 
 \beq
 \label{10dmetric}
 \d s^2 = e^{2 A} \d s^2_{(4)} + \d s^2_{(6)} \, ,
 \eeq
and $F$  is the sum of the RR field strength on  $M$,  $F = F_1 + F_3 + F_5$. 
The ten-dimensional fluxes are defined in terms of $F$ by
\beq
F_{(10)} = {\rm vol}_4 \wedge \lambda (\ast F) + F \, .
\eeq

The complex number $\mu$ determines the size of the AdS$_4$ cosmological constant 
\beq
\Lambda=-|\mu|^2\,.
\eeq

Notice also that,  for AdS vacua, supersymmetry constraints the norms of the two six-dimensional spinors to be equal \cite{Grana:2006kf}
\bea
 |a|^2 = |b|^2 = e^A \, .  
\eea
Only the relative scale between the spinor being relevant, we can always rescale $\eta_+$ in such a way that
\beq
\bar{b} =  a \,,   \qquad \qquad  \frac{b}{a}=  e^{- i\theta} \, . 
 \eeq

It is convenient to introduce
 the rescaled forms
 \bea
\hat{\omega} =  e^{i \theta} \omega \, , \\
\hat{z}=\frac{\bar{\mu}}{|\mu|}z \, , 
 \eea
but for simplicity of notation, we  will drop  the $\hat{~}$ symbols in the rest of the paper.

\vspace{0.3cm}

Plugging the explicit form of  \eqref{Phim} and \eqref{Phip},  into the SUSY variations  \eqref{eqm}-\eqref{eqpb}, one can deduce the 
general conditions for AdS$_4$ $\mathcal{N}=1$ SUSY vacua in terms of   the forms  $z$, $\omega$, $j$ and the fluxes. 
As discussed in \cite{Petrini:2013ika},   \eqref{eqm}  implies
\beq
\kpar = 0 \qquad \mbox{or} \qquad \cos \theta  = 0 \, .
\eeq
The first choice corresponds to a rigid $\SU(2)$ structure, while the second fixes the relative phase of $a$ and $b$.

The supersymmetry conditions for rigid $\SU(2)$ structure were derived in this language in \cite{Petrini:2013ika} and amount to a set of differential
conditions on the $\SU(2)$ forms
\bea
\label{Phimexpf1}
&& \d (e^{3 A - \phi} z ) =  2 |\mu| e^{2 A - \phi}{\omega}_I \,  , \\ 
\label{Phimexpf2}
&& z \w ( \d j - i H + |\mu| e^{-A} \, \bar{z} \w {\omega}_R ) =0 \, ,  \\
\label{Phipaexpf1}
&& \d (e^{2 A - \phi} {\omega}_I ) = 0 \, , \\
\label{Phipaexpf2}
&&  \d( e^{2 A - \phi}  z \w \bar{z} \w  {\omega}_R ) = 2 i  e^{2 A - \phi}   H \w  {\omega}_I \, ,
\eea
plus  equations for the RR fluxes
\bea
\label{1formbapp}
&& \ast F_5 =  3 e^{-A - \phi} \, |\mu|z_I \, , \\
\label{3formbapp}
&& \ast F_3 = -e ^{ -4 A} \, \d (e^{4 A - \phi} {\omega}_R ) +3 e^{-A - \phi}|\mu|z_R \w j  \, , \\
\label{5formbapp}
&&  \ast F_1 =  -i \, \d (2 A - \phi)  z \w \bar{z}  \w {\omega}_I  -   e^{-\phi} H \w {\omega}_R \nn \\
& &  \qquad  \quad +
\frac{1}{2} e^{-A - \phi} |\mu|z_I \w j \w j  \,. 
\eea

\vspace{0.3cm}

With non trivial fluxes, the $\SU(2)$-structure forms are in general not closed. Their differentials can be expanded into representations  of $\SU(2)$, 
the $\SU(2)$ torsion classes (see for instance \cite{Dall'Agata:2003ir,Lust:2004ig, Petrini:2013ika}). There are 20 $\SU(2)$ torsion classes: 
8 complex singlets, 8 complex doublets  and 4 complex triplets. In view of the application to group manifolds, we choose to parametrize the 
torsions in a slightly different way than what was used in  \cite{Dall'Agata:2003ir,Lust:2004ig, Petrini:2013ika}.    The  form $z$  defines an almost product structure on $M$,  
%given  locally by
%\beq
%\label{products}
%R_m^n = z_m \bar{z}^n + \bar{z}_m z^n - \delta_m^n  \,  , \qquad \qquad m,n=1,  \ldots,  6 \, , 
%\eeq 
which induces a (global)  decomposition of the tangent space
\begin{align}
T M = T_2 M \oplus T_4 M \,, \label{decomposition}
\end{align}
where  the subbundle $T_2 M$ is spanned by the real and imaginary parts of the form $z$.  On $T_4 M$ we define the following basis of two-forms
\beq
T^a=\{j, \omega_R, \omega_I, \tilde{j}_1,  \tilde{j}_2,  \tilde{j}_3\}  \qquad a=1, \dots, 6 \, ,
\eeq
where $j$ and $\omega$ define the $\SU(2)$ structure and the $\tilde{j}_a$ form a triplet of anti self-dual two-forms such that\footnote{Given the decomposition (\ref{decomposition}), we  can write the 6-dimensional Hodge star as
$\ast_6(A\wedge B)=\ast_2 A\wedge \ast_4 B$,  where $A\in T_2 M$ and $B \in T_4 M$. One can derive 
\begin{align}
\ast_2 1=&-z_R\wedge z_I & \ast_2 z_R=&z_I & \ast_2 z_I=&-z_R & \ast_2  (z_R\wedge z_I)=&-1\\
\ast_41=&\frac{1}{2}j\wedge j & \ast_4 j,\omega_R, \omega_I=&j,\omega_R, \omega_I & \ast_4 \tilde{j}_i=&-\tilde{j}_i & \ast_4j\wedge j=&2 \, .
\end{align} }

\beq
\tilde{j}_i\wedge\tilde{j}_k= -\delta_{ik}j\wedge j \, , \qquad \qquad 
\tilde{j}_i\wedge j= \tilde{j}_i\wedge \omega =0 \, .  
 \eeq

\vspace{0.4cm}

In what follows, in order to   simplify the computations, we will  set to zero  the torsion classes in the \textbf{2} of $\SU(2)$\footnote{Notice that in presence of orientifolds, these torsion classes are automatically put to zero.} and we will parametrize the exterior derivatives of the two-forms $T^a$ as follows
\bea
&&  \d z= S_0 z_R\wedge z_I+S_a T^a \, , \label{torsions1}\\
&&  \d T^a=(M_1)^a_{~b}z_R\wedge T^b+(M_2)^a_{~b}z_I\wedge T^b \, , \label{torsions2}
\eea 
where  $S_a$ are seven complex scalars and $M_i$ are two real torsion matrices of the  form 
\begin{align}
 M_i=&\left(\begin{array}{cccccc}
        t^1_{i1} & t^1_{i2} & t^1_{i3} & t^1_{i4} & t^1_{i5} & t^1_{i6}\\ 
	-t^1_{i2} & t^1_{i1} & t^2_{i3} & t^2_{i4} & t^2_{i5} & t^2_{i6}\\
	-t^1_{i3} & -t^2_{i3} & t^1_{i1} & t^3_{i4} & t^3_{i5} & t^3_{i6}\\
	t^1_{i4} & t^2_{i4} & t^3_{i4} & t^1_{i1} & t^4_{i5} & t^4_{i6}\\
	t^1_{i5} & t^2_{i5} & t^3_{i5} & -t^4_{i5} & t^1_{i1} & t^5_{i6}\\
	t^1_{i6} & t^2_{i6} & t^3_{i6} & -t^4_{i6} & -t^5_{i6} & t^1_{i1}
       \end{array}
\right) \, .
\end{align}

With an abuse of notation,  we will call torsions  the  singlets $S_a$ and the coefficient of the matrices $M_i$.
The redundancy in the elements $t^a_{ib}$ is fixed by imposing on  the two-forms $T^a$ 
the constraints 
\bea
&& \d (T^a \wedge T^b)=0 \, ,  \nn \\
&& \d^2 T^a = 0 \, . 
\eea
We have already enforced the first set of the above conditions. We still have to impose equations of the type $\d( \d T^a)=0$, which are much harder to solve 
since they  are quadratic in the coefficients $t^i_{aj}$. We will come back to this problem in Section \ref{quadratic} where we found a way to linearly solve them.

\section{Looking for vacua}
\label{looking}

In order to find  $\mathcal{N}=1$ supersymmetric vacua one has to solve the supersymmetry variations and then the Bianchi identities for the fluxes.
In this formalism, the supersymmetry conditions  are easy to solve, since they reduce to a set of linear algebraic constraints for the elements of the matrices $M_i$. The Bianchi identities are harder to handle since they require solving the quadratic constraints $\d^2 T^a$ we introduced above.
In this section we will  solve the supersymmetry equations \eqref{Phimexpf1} - \eqref{Phipaexpf2}. Then we will solve the constraints
$\d^2 T^a$ in the case of group manifolds. Finally we will solve the Bianchi identities: we will show
that there exist no sourceless solutions  and give the list of all group manifolds admitting a vacuum.

\subsection{SUSY constraints}

In order to solve the SUSY variations \eqref{Phimexpf1} - \eqref{Phipaexpf2}, besides  \eqref{torsions1} and  \eqref{torsions2}  for the $\SU(2)$ structure forms,
we need a similar decomposition of  the NS $H$ flux 
\begin{align}\label{Hflux}
 H=h_{1a}z_R \w T^a+h_{2a}z_I\w T^a \, .
\end{align}

Then we use the SUSY equations to put constraints on the torsions classes and on the coefficients of $H$.  Indeed, by plugging 
\eqref{torsions1}, \eqref{torsions2} and \eqref{Hflux} into \eqref{Phimexpf1}-\eqref{Phipaexpf2} (we take  $\phi$ and $A$ constant), one can show that 
\begin{align}
 S_0=&S_1=S_2=S_4=S_5=S_6=0 \, ,  \\
S_3=&2e^{-A}|\mu| \, ,  \\
 h_{11}=&h_{21}=h_{13}=h_{13}=0 \, , \label{susytorsion}
\end{align}
and that the torsion matrices take the following forms :
\begin{align}\label{mattorsion1}
  M_1=&\left(\begin{array}{cccccc}
       0 & -(2e^{-A}|\mu|+h_{22}) & 0 & -h_{24} & -h_{25} & -h_{26}\\ 
	2e^{-A}|\mu|+h_{22} &0 & 0 & t^2_{14} & t^2_{15} & t^2_{16}\\
	0 & 0 & 0 & 0 & 0 &0\\
	-h_{24} & t^2_{14} & 0 &0 & t^4_{15} & t^4_{16}\\
	-h_{25} & t^2_{15} &0 & -t^4_{15} & 0 & t^5_{16}\\
	-h_{26} & t^2_{16} &0 & -t^4_{16} & -t^5_{16} & 0
       \end{array} 
\right) \, , \\
M_2=&\left(\begin{array}{cccccc}
       0 & h_{12} & 0 & h_{14} & h_{15} & h_{16}\\ 
	-h_{12} &0 & 0 & t^2_{24} & t^2_{25} & t^2_{26}\\
	0 & 0 & 0 & 0 & 0 &0\\
	h_{14} & t^2_{24} & 0 &0 & t^4_{25} & t^4_{26}\\
	h_{15} & t^2_{25} &0 & -t^4_{25} & 0 & t^5_{26}\\
	h_{16} & t^2_{26} &0 & -t^4_{26} & -t^5_{26} & 0
       \end{array}\label{mattorsion2}
\right) \, . 
\end{align}

The RR fluxes are then determined from
\eqref{1formbapp}-\eqref{5formbapp}.

\subsection{The quadratic equations for group manifolds}
\label{quadratic}

Before looking at the Bianchi identities, we will now solve the quadratic equations of the form $\d(\d T^a)=0$. Unfortunately,
plugging \eqref{susytorsion}-\eqref{mattorsion2} into  \eqref{torsions2}, the equations $\d(\d T^a)=0$ are quadratic in the torsion coefficients $t^a_{ib}$ and 
the flux parameters  $h_{ia}$, so that it is not an easy task to solve them. We will limit ourselves to group manifolds and enforce these quadratic conditions directly on the basis of one-forms $e^i$, which will simplify the problem.

\vspace{0.3cm}

As already said, we are interested in  six-dimensional homogeneous  group manifolds.  A homogeneous  group manifold is specified 
by a basis  of globally defined one forms,  $e^i$,  satisfying the Maurer-Cartan equations 
\beq
\d e^i=-\frac{1}{2}f^i_{jk}e^j \w e^k\,,
\eeq
with $f^i_{jk}$ constant.  In this case imposing $\d^2 e^i = 0$ gives the Jacobi identities
\beq
f^i_{[j k} f^k_{l] i} = 0 \, .
\eeq

By an appropriate choice of basis $e^i$,   the $\SU(2)$-structure can always be put in the form
\begin{align}
 z=& z_1e^1+i z_2e^2\, ,\nonumber\\
j=&j_1e^{36}+j_2 e^{45}\, ,\nonumber\\
\omega_R=&\frac{j_1j_2}{\omega_{1}}e^{34} + \omega_{1}e^{56}\, ,\nonumber\\
\omega_I=&-\frac{j_1j_2}{\omega_{2}}e^{35}+\omega_{2}e^{46}\,  ,\label{ansatzform1} 
\end{align}
where  all the coefficient are real.  Similarly, the  $\SU(2)$ triplet of anti-self dual two forms $\tilde{j}_i$ can be written as 
\begin{align}
\tilde{j}_1 =& j_1e^{36} - j_2 e^{45}  \, ,\nonumber \\
\tilde{j}_2  =& -  \frac{j_1j_2}{\omega_{1}}e^{34} + \omega_{1}e^{56}  \, ,\nonumber \\
\tilde{j}_3  =& -\frac{j_1j_2}{\omega_{2}}e^{35} - \omega_{2}e^{46} \, .  \label{ansatzform2}
\end{align}

\subsubsection{Solving the Jacobi Identities on group manifold}\label{Jacobi}
\label{jacobi}

Once we use the expressions  \eqref{ansatzform1} and \eqref{ansatzform2} for  $z$ and $T^a$,  the quadratic constraints $\d^2 T^a=0$ is clearly 
satisfied since we know that  the one-forms $e^i$ must satisfy the Jacobi identity $\d^2 e^i = 0$. 
The idea is then to express the structure constants in terms of the torsion parameters and the fluxes $h_{ia}$ and then solve explicitly the Jacobi
identities in terms of such parameters.

\vspace{0.3cm}

Plugging  \eqref{ansatzform1} and \eqref{ansatzform2}  in the torsion equations \eqref{torsions1} and \eqref{torsions2} gives the structure
constant in terms  of the  torsion  $t^a_{ib}$ and  the coefficient $h_{ia}$. Since they are not particularly enlightening we give the explicit expressions
in Appendix \ref{structureconstants}. 
Then we can impose the  Jacobi identities.  Fortunately some of these equations are linear in $t^a_{ib}$ and in $h_{ia}$ after we impose the SUSY constraints and one can solve the full set  of the Jacobi identities. The result is quite simple. It puts all coefficients of the matrix \eqref{mattorsion1} to zero whereas \eqref{mattorsion2} is unchanged. Concerning the structure constants, here it is also very simple, $f^3_{j1},~f^4_{j1},~f^5_{j1},~f^6_{j1}$ are put to zero and the other ones are left unchanged.

\subsubsection{Intermediate $\SU(2)$ structures on group manifolds}

Up to know we restricted our analysis to the case of rigid  $\SU(2)$ structure $\kpar =0$. 
In this section we will briefly discuss the case of intermediate $\SU(2)$ structure when restricted on group manifolds.
As we saw in Section \ref{susysect}, having both $\kpar \neq 0$ and $\kper \neq 0$ forces to set 
 $\theta = - \frac{\pi}{2}$. Then, for constant dilaton and warp factor, 
\eqref{eqm} and \eqref{eqp} give 
\bea
&& \d z_R = -\frac{2 e^{-A} |\mu|}{\kper} \left[ \kper \omega_R+\kpar (j+z_R\wedge z_I) \right] \, ,\\
&& \d z_I= 0 \, , \\
&& \d \omega_R= \frac{\kpar}{\kper^2}\left[-\kper \d j+2 |\mu| e^{-A} z_I\wedge(\kpar j+\kper \omega_R)\right]\\
& &z\wedge \Big[ \d j-i \kper^2 H-i \kper\kpar \d \omega_I - |\mu| e^{-A} \bar{z}\wedge\big(\frac{\kpar}{\kper}(1-2\kper^2)2 i j- \omega_I + 2i \kpar^2\omega_R\big) \Big] = 0 \, , \\
&&i \kper z\wedge \bar{z}\wedge \d\omega_I - 2 \kpar j\wedge H - 2\kper \omega_R \wedge H - i \kpar z \wedge \bar{z}\wedge H = 0 \, ,
\eea
while \eqref{eqpb} can be used,  as always, to determine the RR fluxes
\bea
&&e^{\phi}\ast F_5=3z_I|\mu e^{-A}|\kper \, ,\\
&&e^{\phi}\ast F_3=\kpar H-\kper d\omega_I+3|\mu e^{-A}|\left(\kper z_R \wedge j +\kpar(z_I\wedge\omega_I-z_R\wedge \omega_R)\right) \, ,\\
&&e^{\phi}\ast F_1=\frac{-1}{2\kper}\left(2\kpar \kper j\wedge dj+2\kper^2 \omega_I\wedge H-|\mu e^{-A}|z_I\wedge j\wedge j(1+3\kpar^2)\right) \, .
\eea

Repeating the analysis of the SUSY conditions and Jacobi identities in this case, it is lengthy but straightforward to show 
it is not possible to solve these equations.  This  means the only way to have AdS$_4$ vacua with $\mathcal{N}=1$  
is to be on a strict $\SU(2)$-structure manifold.

\subsection{Bianchi identities}

In order to find a supersymmetric vacuum we still have to solve the Bianchi identities for the NS and RR fluxes. These are the equations
that tell us whether we can have sourceless solutions and, if sources are needed, of which type they must be.  In this work we assume that there
are no NS5 branes, meaning  that the NS Bianchi identity, $\d H=0$, simply puts more constraints on the parameters $t^a_{ib}$ and $h_{ia}$.
Such constraints are in general quadratic in  $t^a_{ib}$ and $h_{ia}$ and admit different possible solutions.  We give the explicit form of
the NS Bianchi identities as well as its various classes of solutions in Appendix \ref{NSBI}.

Since one of the aim of this paper is to see whether sourceless solutions are possible, we discuss in detail the RR Bianchi identities
\bea
&& \d F_1=  \delta(D7/O7) \, ,\\
&& \d F_3-H\w F_1 =  \delta(D5/O5) \, ,\\
&& \d F_5-H \w F_3=  \delta(D3/O3) \, ,
\eea
where  $\delta(Dp/Op)$ is  the charge densities of the space-filling sources.  It is easy to see that  for our solutions 
\beq
\delta(D3/O3)  =0 \, . 
\eeq
The other two equations give 
\bea
\label{7ch}
 \delta(D7/O7) &=& -10e^{-\phi} |\mu e^{-A}|^2 \omega_I \, , \\
 \label{5ch}
 \delta(D5/O5) &=& 
 e^{-\phi}z_R \w z_I \w \left[(10|\mu e^{-A}|^2+3h_{12}^2+(t^2_{24})^2+(t^2_{24})^2+(t^2_{25})^2+(t^2_{26})^2)\omega_R\right. \nonumber\\ 
&& \left.+(h_{14}t^2_{24}+h_{15}t^2_{25}+h_{16}t^2_{26})j+(3h_{12}h_{14}-t^2_{25}t^4_{25}-t^2_{26}t^4_{26})\tilde{j}_1\right. \nonumber\\ 
&&   \left.+(3h_{12}h_{15}+t^2_{24}t^4_{25}-t^2_{26}t^5_{26})\tilde{j}_2+(3h_{12}h_{16}+t^2_{24}t^4_{26}+t^2_{25}t^5_{26})\tilde{j}_3\right]\, .
\eea

Note that we haven't enforced the condition $\d H=0$ in above expressions  above.  A common feature of the solutions of Appendix \ref{NSBI}
of the NS Bianchi identities is that the coefficient in front of $z_R \w z_I\w j$ in $\delta(D5/O5)$ vanishes.

\subsubsection{Sourceless solutions}

For constant warp factor, there are some general arguments allowing to characterize the manifolds that could give  sourceless solutions.
In particular, by taking appropriate linear combinations of the 10-dimensional equations of motion in absence of sources, we arrive at 
\begin{align}
& R_4=2V = -2\sum_p  |F_p|^2 < 0\,,\\
& R_6 = \sum_p \frac{9-p}{2} |F_p|^2>0\, , 
\end{align}
which imply that in order to have sourceless solutions the internal manifold must have positive curvature. This condition rules out 
all nilmanifolds as possible candidates for sourceless solutions, but still leaves open the possibility of having other group manifolds.

However, from the Bianchi identities in the previous section, we immediately see that it is not possible to avoid O7 or D7 sources
(see \eqref{7ch}), thus 
ruling  out the possibility of having sourceless AdS SUSY vacua in type IIB. This is very unlike type IIA where many source free solutions where found \cite{Tomasiello:2007eq,Koerber:2010rn}. 

Since T-duality does not create sources out of nothing, one might wonder whether it is possible to generate sourceless solutions in
IIB  T-dualizing one of the known sourceless solutions of IIA, for instance those in \cite{Tomasiello:2007eq}.  One can check that  in all known
IIA solutions the isometries on which one could T-dualize correspond to cycles that shrink to zero at some points, thus giving rise to non compact solutions, which are outside the classification of this paper.

\subsubsection{Sign of charges}

In this section we want to  look at the sign of the charges in the Bianchi identities 
 in order to see if D-branes or O-planes are involved. We will follow  the approach of  \cite{Grana:2006kf}.
The source terms \eqref{7ch} and \eqref{5ch} can be written as sum of  decomposable forms (this is necessary if they are to describe
a plane) 
\begin{align*}
 \delta(D7/O7)=&\sum{} N_{(D7/O7)ij} e^i\wedge e^j \, , \\ 
 \delta(D5/O5)=&\sum{} N_{(D5/O5)ijkl} e^i\wedge e^j\wedge e^k\wedge e^l \, .
\end{align*}
We want to rewrite them in terms of the volume forms orthogonal to the cycles wrapped by the source. We  define $\mathrm{vol}^{ij}$  as the decomposable form proportional to $e^i\wedge e^j$ and normalized such that 
\beq
\langle \mathrm{vol}^{ij},\Im \Phi_+ \rangle = - \mathrm{vol}_6 =-  i \langle \Phi_+,\overline{\Phi}_+ \rangle  \, . 
\eeq
Similarly, one can introduce  $\mathrm{vol}^{ijkl}$. With these definitions, the sources can be written as
\begin{align}
\label{chargeden}
 \delta(D7/O7)=&\sum{} n_{(D7/O7)ij} \mathrm{vol}^{ij} \, , \\
  \delta(D5/O5)=&\sum{} n_{(D5/O5)ijkl} \mathrm{vol}^{ijkl} \, , 
\end{align}
where $n_{(D7/O7)ij}$ and $ n_{(D5/O5)ijkl}$ can be seen as  densities of charges on the cycle wrapped by the sources, they give the sign of the charges. On the other hand, $N_{(D7/O7)ij}$ and $ N_{(D5/O5)ijkl}$ are more similar to total charges and these should be generally of order one on a solution since they are directly related to the number of D-branes or orientifolds.
\vspace{0.3cm} 

In order to fix our conventions for the signs of $n_{(D7/O7)ij}$ and $ n_{(D5/O5)ijkl}$ we recall briefly  the case of Minkowski compactifications.
In this case, one can show that the tadpole condition is equivalent to \cite{Grana:2006kf}
\bea
\sum_p |F_p|^2  \, {\rm vol}_6  &=&  \langle F ,  \ast \lambda( F)  \rangle  \nonumber \\
&=&  \sum_i n_i \langle \mathrm{vol}^{i},\Im \Phi_+ \rangle = -  \sum_i n_i \, {\rm vol}_6 \, , 
\eea
where, for simplicity, we denoted by $n_i$ the charge density of a generic source and by ${\rm vol}^i$ the corresponding transverse volume.
In order to solve the tadpole condition, we need a net orientifold charge. This means that
we can associate orientifolds to 
a negative charge density $n_i$  and branes to a positive one.   
In this paper we will use the same conventions: if  $n_{(D7/O7)ij}$ and/or  $ n_{(D5/O5)ijkl}$ are negative we have an overall O-plane
charge and viceversa.

\begin{comment}
There is an important caveat to this. We can't define $\mathrm{vol}^{ij}$ and $\mathrm{vol}^{ijkl}$ for all $ij$ and all $ijkl$. Indeed, with our ansatz 
\eqref{ansatzform1} and \eqref{ansatzform2}  one can only define $\mathrm{vol}^{35}$, $\mathrm{vol}^{46}$, $\mathrm{vol}^{1234}$ and $\mathrm{vol}^{1256}$. As we will see, orthogonal to these directions, there will be presence of orientifold. Then, the constraints coming from the SUSY equations, the structure of the manifold and the presence of these orientifolds will be sufficient to force  $\delta(D7/O7)$ and $ \delta(D5/O5)$ to be in these four directions.
\end{comment}

\vspace{0.3cm} 

Using the definition above, one finds that 
\begin{align}
n_{(D7/O7)35}=n_{(D7/O7)46}=-\frac{5e^{-3A-\phi}}{4}|\mu|^2<0\,,
\end{align}
implying that,  in order to have an AdS SUSY solution on a group manifold, one must  always have two intersecting O7-planes. 
Since we have intersecting orientifolds, we will only consider smeared sources. 
 The D5/O5 case does not give any general condition, and we analyze it case by case in the next section.

\section{Scanning  group Manifolds}
\label{scanning}

We can use the formalism developed above to perform a complete search of group manifolds admitting  $\mathcal{N}=1$ AdS$_4$ solutions in type IIB.
This amounts to determining the set of structure constants that solve both the supersymmetry conditions and the Bianchi identities.

\vspace{0.3cm}

The logic is the following.  We first impose the supersymmetry constraints \eqref{Phimexpf1}-\eqref{5formbapp} and the Jacobi identities for the $\SU(2)$ structure on the generic group manifold.% This gives  the relations in Section \ref{jacobi} among some of the torsions and $\SU(2)$ structure moduli in terms of the structure constants.

We then look at the Bianchi identities: first the RR 1-form, then the NS 3-form and finally the RR 3-form.
 The BI for the RR 1-form implies that the solutions necessarily have intersecting O7-planes. Then the compatibility of the algebrae with the orientifold involution forces to set 
\beq
h_{16}= t^4_{26}= t^2_{26}= t^5_{26}=0 \, . 
\eeq

As given in  Appendix \ref{structureconstants}, the expressions for $f^i_{jk}$ are not very manageable. Moreover, in looking for solutions, one usually proceed in the opposite way: 
given an internal manifold, one would like to find the coefficients of the torsions and $H$ flux in terms of the structure constants. 
By inverting these relation we find the following expressions for the parameters in the $\SU(2)$
\beq
 \omega_2=-\frac{f^1_{46}z_1}{2e^{-A}|\mu|}  \qquad j_2=-\frac{D}{4j_1} \, , \\
 \eeq
with $D=\frac{f^1_{35}f^1_{46}z_1^2}{4|e^{-A}\mu|^2}$, the torsions 
\begin{align}
t^4_{25}=&-\frac{1}{8j_1z_2\omega_1}\left(-f^4_{26}D-4f^5_{23}\omega_1^2+4j_1^2(f^3_{25}-\frac{4f^6_{24}\omega_1^2}{D})\right) \, , \nn \\
 t^2_{25}=&\frac{f^3_{23}+f^4_{24}}{z_2} \, , \nn \\
t^2_{24}=&-\frac{1}{8j_1z_2\omega_1}\left(-f^4_{26}D+4f^5_{23}\omega_1^2+4j_1^2(f^3_{25}-\frac{4f^6_{24}\omega_1^2}{D})\right) \, ,
\end{align}
and the $H$ flux
\begin{align}
h_{15}=&-\frac{1}{8j_1z_2\omega_1}\left(f^4_{26}D+4f^5_{23}\omega_1^2+4j_1^2(f^3_{25}+\frac{4f^6_{24}\omega_1^2}{D})\right) \, ,\nn \\
h_{12}=&-\frac{1}{8j_1z_2\omega_1}\left( f^4_{26}D-4f^5_{23}\omega_1^2+4j_1^2(f^3_{25}-4\frac{f^6_{24}\omega_1^2}{D})\right) \,,  \nn  \\
h_{14}=& \frac{-f^3_{23}+f^4_{24}}{z_2} \, .
\end{align}
 We also find  that 
\beq
f^5_{25}=-f^3_{23} \qquad \qquad f^6_{26}=-f^4_{24} \, . 
\eeq
The non-zero free parameters are now : $j_1$, $z_1$,$z_2$, $\omega_1$, $f^1_{35}$, $f^1_{46}$ and $f^3_{23}$, $f^3_{25}$, $f^4_{24}$, $f^4_{26}$,
$f^5_{23}$, $f^6_{24}$.

\vspace{0.3cm}

As already mentioned, the NS Bianchi identity is quadratic in the free parameters. As a result we find three different solutions, which we
give in detail in Appendix \ref{NSBI}.

We then study the O5 source equations in each of the three solutions of the NS Bianchi identity. If the charge densities 
$n_{(D5/O5)}$  are negative,  there are  additional constraints on the structure constants due to the presence of an O5-plane in the 
directions $34$ and $56$ 
\beq
f^3_{23}=  f^4_{24}=0 \, . 
\eeq

As a result of this procedure we find that $\mathcal{N}=1$ AdS$_4$  solutions are very rare and all have intersecting O5-planes besides 
the O7 we already mentioned. We give the list of solutions in Table  \ref{tablelistmanifold}
\begin{table}[h]
\begin{center}
\begin{tabular}{|c|c|c|c|}\hline
$\mathrm{algebra}$ &$ O5$ & $\mathrm{type~of~manifold}$ & name\\\hline\hline
$(35+ \epsilon 46,0,0,0,23,24)$ & $\checkmark$ &nilmanifold &$ n3.13,~n3.14$\\ \hline
$(35+ 46,0,0,0,23,0)$ &$ \checkmark$&nilmanifold & $n4.1$\\ \hline
$(35+ 46,0,0,0,0,0)$ &$ \checkmark$&nilmanifold & $n5.1$\\ \hline
$(35+ \epsilon 46,0,25,-\epsilon 26,-\epsilon23,24)$  & $ \checkmark$&solvmanifold & $\mathfrak{g}6.88$ and solv$_1$ \\ \hline
\end{tabular}
\caption{list of group manifolds admitting an AdS SUSY solution ($\epsilon=\pm 1$).}
\label{tablelistmanifold}
\end{center}
\end{table}
\noindent  where the O5 column tells us if there is presence or absence of O5 plane. 

The solutions on nilmanifolds  have already been studied in this framework in \cite{Petrini:2013ika}, and we refer to that reference for 
details. In the next section we describe the last solution in the table, which, to our knowledge, is new. Concerning the names of the new manifold, look at appendix \ref{solvname} for an detailed explanation\footnote{The author would like to thank David Andriot for a useful discussion about solvmanifold algebras}.

\subsection{A new solution}
\label{newsolution}

As far as the author knows, $(35+ \epsilon 46,0,25,-\epsilon 26,-\epsilon23,24)$ is a new solution in the literature on the subject. But as we will see in appendix \ref{Tduality}, it can be obtained as a T-dual of a Lust-Tsimpis type solution in IIA given in \cite{Grana:2006kf}. 

Before giving the explicit vacua, let's give some properties of the algebra. It is easy to see that it is a solvable algebra. From its   
Killing form 
\begin{align}
\left(\begin{array}{cccccc}
0 & 0 & 0 & 0 & 0 & 0\\
0 & -4 \epsilon & 0 & 0 & 0 & 0\\
0 & 0 & 0 & 0 & 0 & 0\\
0 & 0 & 0 & 0 & 0 & 0\\
0 & 0 & 0 & 0 & 0 & 0\\
0 & 0 & 0 & 0 & 0 & 0
\end{array}\right) \, , 
\end{align}
one can see that this algebra is compact if and only if $\epsilon=-1$.

 We give here all the quantities of interest on the vacua.
The $\SU(2)$ structure forms are\footnote{Similarly  the triplet of anti-self dual two-forms
are given by
\begin{align}
\tilde{j}_1=&j_1(e^3\wedge e^6+\epsilon e^4\wedge e^5) \, , \nn \\
\tilde{j}_2=&\epsilon \frac{j_1^2}{\omega_1}e^3\wedge e^4+\omega_1 e^5\wedge e^6 \, , \nn \\
\tilde{j}_3=&\epsilon_1 j_1(\epsilon e^3\wedge e^5-e^4 \wedge e^6) \, . 
\end{align}}
\begin{align}
z=&\epsilon \epsilon_1 2|e^{-A}\mu|j_1 e^1+z_2 e^2  \, , \nn \\
j=&j_1(e^3\wedge e^6-\epsilon e^4\wedge e^5) \, , \nn \\
\omega_R=&-\epsilon \frac{j_1^2}{\omega_1}e^3\wedge e^4+\omega_1 e^5\wedge e^6 \, , \nn \\
\omega_I=&\epsilon_1 j_1(\epsilon e^3\wedge e^5+e^4 \wedge e^6) \, .
\end{align}
We can easily compute the Ricci scalar 
\begin{align}
 R_6=&-4|\mu e^{-A}|^2-\left(\frac{j_1^2-\epsilon \omega_1^2}{j_1 z_2 \omega_1}\right)^2 \, .
\end{align} 
The solution has non trivial NS and RR fluxes
\begin{align}
H=&\frac{1}{\omega_1}2|\mu e^{-A}|z_2(\epsilon j_1^2 e^2\wedge e^3\wedge e^4-\omega_1e^2\wedge e^5\wedge e^6) \, , \\
F_1=&10e^{-\phi}\epsilon_1 |e^{-A}\mu|^2j_1e^1 \, , \\
F_3=&e^{-\phi}\frac{|e^{-A}\mu|j_1}{z_2 \omega_1}\left[3z_2^2 \omega_1^2(\epsilon e^2\wedge e^4\wedge e^5-e^2 \wedge e^3\wedge e^6)\right.\nonumber\\
&\hspace{2.5cm}\left.+2\epsilon_1( j_1^2-\epsilon\omega_1^2)(e^1 \wedge e^3 \wedge e^6+ \epsilon_1 e^1\wedge e^4 \wedge e^5)\right] \, , \\
F_5=&-6e^{-\phi}\epsilon\epsilon_1 |e^{-A}\mu|^2j_1^3 e^1 \wedge e^3 \wedge e^4 \wedge e^5 \wedge e^6 \, , 
\end{align}
giving rise to the the charges 
\begin{comment}
\begin{align}
 N_{(D7/O7)35}=&10e^{-\phi}\epsilon_1|\mu|^2 j_1\\
N_{(D7/O7)46}=&-10e^{-\phi}\epsilon\epsilon_1|\mu|^2 j_1\\
N_{(D5/O5)1234}=&\frac{\epsilon \epsilon_1 4e^{-\phi}|\mu|j_1}{z_2\omega_1}\left(j_1^2(-\epsilon+5|\mu|z_2^2)+\omega_1^2\right)\\
N_{(D5/O5)1256}=&-\frac{\epsilon_1 4e^{-\phi}|\mu|j_1}{z_2\omega_1}\left(j_1^2+(-\epsilon+5|\mu|z_2^2)\omega_1^2\right)
\end{align}
\end{comment}
\begin{align}
n_{(D5/O5)1234}=&\frac{e^{-\phi}}{4}\left(-5|e^{-A}\mu|^2+\frac{\epsilon-\frac{\omega_1^2}{j_1^2}}{z_2^2}\right) \, , \\
n_{(D5/O5)1256}=&\frac{e^{-\phi}}{4}\left(-5|e^{-A}\mu|^2+\frac{\epsilon-\frac{j_1^2}{\omega_1^2}}{z_2^2}\right) \, . 
\end{align}
\noindent with $\epsilon_1=\pm 1$. Notice that we still have three free parameters ($j_1$, $z_2$ and $\omega_1$).  Note also that 
\begin{align}
n_{(D5/O5)1234}+n_{(D5/O5)1256}=-\frac{5e^{-\phi}}{2}|e^{-A}\mu|^2-e^{-\phi}\left(\frac{j_1}{2z_2\omega_1}-\frac{\epsilon \omega_1}{2 j_1 z_2}\right)^2<0\,,
\end{align}
\noindent which means that at least one of them is negative and so there is, as we said, at least one O5-plane.

\vspace{0.3cm}

A natural question to ask is whether this solution has a good classical supergravity limit, it allows for separation of scales
and  large volume limit. Using the same approach of  \cite{Petrini:2013ika} we obtain
%We also give the study of separation of scale as done in \cite{Petrini:2013ika} :
\begin{center}
  \begin{tabular}{| c | c | c | c| }
    \hline
   weak coupling & Large volume & scale separation (1) & scale separation (2)\\ \hline\hline
    \checkmark           & $\times$     &  \checkmark          &  \checkmark         \\ \hline
  \end{tabular}
\end{center}

\section{Conclusion}

In this paper,  we studied AdS$_4$ flux vacua with $\mathcal{N}=1$ SUSY  in type IIB supergravity with and without sources.
It is well known that such vacua are only possible on manifold with $\SU(2)$ structure.  
The presence of fluxes implies that the intrinsic torsions are non trivial.  
For constant dilaton and warp factor, it was shown in \cite{Lust:2004ig} that, if only the singlets in the torsions are non-zero,  no supersymmetric  sourceless solutions exist.  
In this  paper we extend the analysis to a more general class of $\SU(2)$ torsions where only the vector representations are set to zero, but we keep the warp factor and dilaton constant.
In order to be able to solve the Bianchi identities, we have  also restrict our analysis to group manifold. 
Under this restriction we show that, contrary to type IIA, no sourceless supersymmetric solutions are allowed.
In particular we show that O7 and O5/D5 sources are necessarily present.

We also performed an exhaustive scan of possible $\mathcal{N}=1$ AdS$_4$  vacua on group manifolds. Such solutions are quite rare. The presence of 
the orientifold planes and supersymmetry severely constraint the form of the structure constant and we find that very few solutions are possible.
Namely some nilmanifolds that had already been studied in \cite{Petrini:2013ika} and one new solution on a solvmanifold.

Our analysis assumes constant dilaton and warp factor and also does not cover the most general $\SU(2)$ structure manifold, since we
set to zero all the torsions in vector representations.  Notice that such torsions  are put automatically to zero in presence of orientifold,  so one can hope that by turning them on, one can evade the necessity of O-planes. One can also devise a method to solve the quadratic equations in another way that the one used here in order to free oneself from the use of group manifold only. The general method would still work and one could discover new vacua.

There are also more general developments one can do that are not directly related to the method used. For example, one should understand how one could have fully localized sources and the use of the pure spinors formalism could help \cite{Saracco:2012wc}. As we said we can try to look to non SUSY vacua close to the SUSY one and hope to find stable dS vacua (unstable ones were found for example in \cite{Danielsson:2009ff,Danielsson:2010bc,Andriot:2010ju,Danielsson:2011au,Caviezel:2008tf,Flauger:2008ad}). But with more vacua available, there is hope to find a stable dS vacuum.

\vspace{0.5cm}
\textbf{Acknowledgements}
\vspace{0.5cm}

I would like to thank Michela Petrini, Alessandro Tomasiello and Thomas Van Riet for helpful discussions and advice.

\appendix

\section{Structure constants}\label{structureconstants}

Here we give the expression of the structure constants in term of torsion classes and coefficients of the NS three form $H$ after having solved the SUSY equations. We only list the non-zero structure constants  
\begin{align*}
f^1_{jk}=&\left(
\begin{array}{cccccc}
 0 & 0 & 0 & 0 & 0 & 0 \\
 0 & 0 & 0 & 0 & 0 & 0 \\
 0 & 0 & 0 & 0 & \frac{|\mu|e^{-A} j_1 j_2}{z_1 \omega_2} & 0 \\
 0 & 0 & 0 & 0 & 0 & -\frac{|\mu|e^{-A} \omega_2}{z_1} \\
 0 & 0 & -\frac{|\mu|e^{-A} j_1 j_2}{z_1 \omega_2} & 0 & 0 & 0 \\
 0 & 0 & 0 & \frac{|\mu|e^{-A} \omega_2}{z_1} & 0 & 0 \\
\end{array}
\right) & 
f^2_{jk}=&\left(
\begin{array}{cccccc}
 0 & 0 & 0 & 0 & 0 & 0 \\
 0 & 0 & 0 & 0 & 0 & 0 \\
 0 & 0 & 0 & 0 & 0 & 0 \\
 0 & 0 & 0 & 0 & 0 & 0 \\
 0 & 0 & 0 & 0 & 0 & 0 \\
 0 & 0 & 0 & 0 & 0 & 0 \\
\end{array}
\right)
\end{align*}
\begin{align*}
f^3_{j1}=&\left(
\begin{array}{c}
 0  \\
 0  \\
 -\frac{1}{4} (h_{24}+t^2_{15}) z_1  \\
\frac{(h_{26}-t^4_{16}) z_1 \omega_2}{4 j_1}\\
 -\frac{\left(h_{22}+h_{25}-t^2_{14}-t^4_{15}+2 |\mu|e^{-A}\right) z_1 \omega_1}{4 j_1}\\
 \frac{(t^2_{16}-t^5_{16}) z_1 \omega_2 \omega_1}{4 j_1 j_2}  \\
\end{array}
\right) & 
f^3_{j2}=&\left(
\begin{array}{c}
  0 \\
 0  \\
  \frac{1}{4} (h_{14}-t^2_{25}) z_2 \\
 -\frac{(h_{16}+t^4_{26}) z_2 \omega_2}{4 j_1} \\
  \frac{(h_{12}+h_{15}+t^2_{24}+t^4_{25}) z_2 \omega_1}{4 j_1} \\
  \frac{(t^2_{26}-t^5_{26}) z_2 \omega_2 \omega_1}{4 j_1 j_2} \\
\end{array}
\right)
\end{align*}
\begin{align*}
f^4_{j1}=&\left(
\begin{array}{c}
 0 \\
 0 \\
 \frac{(h_{26}+t^4_{16}) j_1 z_1}{4 \omega_2} \\
 \frac{1}{4} (h_{24}-t^2_{15}) z_1 \\
 \frac{(t^5_{16}-t^2_{16}) z_1 \omega_1}{4 \omega_2} \\
 \frac{\left(h_{22}+h_{25}+t^2_{14}+t^4_{15}+2 |\mu|e^{-A}\right) z_1 \omega_1}{4 j_2}  \\
\end{array}
\right) & 
f^4_{j2}=&\left(
\begin{array}{c}
0  \\
  0  \\
  \frac{(t^4_{26}-h_{16}) j_1 z_2}{4 \omega_2}\\
 -\frac{1}{4} (h_{14}+t^2_{25}) z_2 \\
  \frac{(t^5_{26}-t^2_{26}) z_2 \omega_1}{4 \omega_2}\\
  \frac{(-h_{12}-h_{15}+t^2_{24}+t^4_{25}) z_2 \omega_1}{4 j_2}\\
\end{array}
\right)
\end{align*}
\begin{align*}
f^5_{j1}=&\left(
\begin{array}{c}
 0 \\
 0 \\
 \frac{\left(h_{22}-h_{25}+t^2_{14}-t^4_{15}+2 |\mu|e^{-A}\right) j_1 z_1}{4 \omega_1}  \\
 -\frac{(t^2_{16}+t^5_{16}) z_1 \omega_2}{4 \omega_1}  \\
 \frac{1}{4} (h_{24}+t^2_{15}) z_1 \\
 \frac{(h_{26}+t^4_{16}) z_1 \omega_2}{4 j_2} \\
\end{array}
\right) & 
f^5_{j2}=&\left(
\begin{array}{c}
 0 \\
 0\\
  \frac{(-h_{12}+h_{15}+t^2_{24}-t^4_{25}) j_1 z_2}{4 \omega_1} \\
-\frac{(t^2_{26}+t^5_{26}) z_2 \omega_2}{4 \omega_1}  \\
 \frac{1}{4} (t^2_{25}-h_{14}) z_2 \\
 \frac{(t^4_{26}-h_{16}) z_2 \omega_2}{4 j_2}\\
\end{array}
\right)
\end{align*}
\begin{align*}
f^6_{j1}=&\left(
\begin{array}{c}
 0 \\
 0  \\
 \frac{(t^2_{16}+t^5_{16}) j_1 j_2 z_1}{4 \omega_2 \omega_1}\\
 -\frac{\left(h_{22}-h_{25}-t^2_{14}+t^4_{15}+2 |\mu|e^{-A}\right) j_2 z_1}{4 \omega_1} \\
 \frac{(h_{26}-t^4_{16}) j_2 z_1}{4 \omega_2}  \\
 \frac{1}{4} (t^2_{15}-h_{24}) z_1  \\
\end{array}
\right) &
f^6_{j2}=&\left(
\begin{array}{c}
 0  \\
 0  \\
  \frac{(t^2_{26}+t^5_{26}) j_1 j_2 z_2}{4 \omega_2 \omega_1}  \\
  \frac{(h_{12}-h_{15}+t^2_{24}-t^4_{25}) j_2 z_2}{4 \omega_1}  \\
  -\frac{(h_{16}+t^4_{26}) j_2 z_2}{4 \omega_2} \\
  \frac{1}{4} (h_{14}+t^2_{25}) z_2  \\
\end{array}
\right)
\end{align*}

\section{Solving the NS Bianchi Identity}
\label{NSBI}
In this appendix we list the explicit solutions of the NS Bianchi identity
\begin{align}
dH=&(h_{12}^2-h_{14}^2-h_{15}^2)z_R\wedge z_I \wedge j-(h_{14}t^2_{24}+h_{15}t^2_{25})z_R\wedge z_I \wedge \omega_R\nonumber\\
+&(h_{15}t^4_{25}-h_{12}t^2_{24})z_R \wedge z_I \wedge \tilde{j}_1-(h_{12}t^2_{25}+h_{14}t^4_{25})z_R \wedge z_I \wedge \tilde{j}_2\\
=&0\nonumber
\end{align}
We found three different types of solutions (here $\epsilon=\pm 1$)
\begin{itemize}
\item[] Type 1
\begin{align}
h_{12}=&\epsilon\sqrt{h_{14}^2+h_{15}^2} & t^2_{25}=&-\frac{h_{14}t^2_{24}}{h_{15}} & t^4_{25}=&\epsilon\frac{\sqrt{h_{14}^2+h_{15}^2}t^2_{24}}{h_{15}}
\end{align}
\item[] Type 2
\begin{align}
h_{12}=&\epsilon h_{14} & h_{15}=&0 & t^2_{24}=&0 & t^4_{25}=&-\epsilon t^2_{25} 
\end{align}
\item[] Type 3
\begin{align}
h_{12}=&0 & h_{14}=&0 & h_{15}=&0
\end{align}
\end{itemize}

\section{T-duality of (0,0,25,-26,-23,24)}\label{Tduality}

In type IIA theories, AdS$_4$ vacua with $\mathcal{N}=1$ supersymmetry are only possible on $\SU(3)$ structure manifolds.
The general class of solutions was derived in \cite{Lust:2004ig}, and later in the language of Generalized Geometry in \cite{Grana:2006kf}.
In \cite{Grana:2006kf}, the authors also give an explicit solution of this type on a solvmanifold with intersecting O6 planes. 
We will first rewrite it in our notation and then do a T-duality in order to recover the  solution of Section \ref{newsolution}. 

The solution is found on the solvmanifold defined by the algebra (0,0,25,-26,-23,24).
The $\SU(3)$ structure has the form  
\begin{align}
 J=&e^{1}\wedge e^2-e^{4}\wedge e^5+e^{3}\wedge e^6 \, , \nn \\
\Omega=&-i(e^1-ie^2)\wedge (e^3+ie^6)\wedge (e^4-i e^5) \, , 
\end{align}
and the non-zero fluxes are 
\begin{align}
H=&2|\mu| \Omega_R \,, \nn \\
F_0=&5|\mu| \, , \nn \\
F_4=&3/2|\mu| J\wedge J \, .
\end{align}
\begin{comment}
which verifies :
\bea
&&(\d + H  \w )(e^{2A - \phi} \Phi_+)  =  2 \mu e^{A-\phi} {\rm Re } \Phi_-,  \\
&&  (\d + H \w )(e^{A -\phi } {\rm Re} \, \Phi_-) =0 \, , \\
&& (\d + H \w )(e^{3 A -\phi } {\rm Im} \, \Phi_-) = 3e^{A-\phi}{\rm Im} \, (\bar{\mu}\Phi_-) +\frac{1}{8} e^{4 A} \ast F \,, 
\eea
 where $\phi$ is the dilaton, $A$ the warp factor, $\Phi_-=\Omega$ and $\Phi_+=e^{-iJ}$. These are the equivalent of \ref{eqm}, \ref{eqp}, \ref{eqpb} in IIA. 
\end{comment}
We can also define the flux part of $H$ with :
\begin{align}
 B=&-2|\mu|(-1+\alpha)e^{4}\wedge e^5\, ,\\
 H_{fl}=&2|\mu|\left(-e^{1}\wedge e^3\wedge e^5-e^{1}\wedge e^4\wedge e^6+2\alpha e^{2}\wedge e^3\wedge e^4-(2-\alpha)e^{2}\wedge e^5\wedge e^6\right)\, ,
\end{align}
with $\alpha$ an arbitrary real parameter. 
%On can see that the directions $1$ and $2$ are isometries. 
The manifold is  $S^1_{\{1\}}
\times M_5$, where $M^5$ is a $T^2_{\{3,5\}}\times T^2_{\{4,6\}}$--fibration over $S^1_{\{2\}}$. Thus we can perform a  T-duality 
along the direction $1$. We will follow the rules for T-duality given in \cite{Grana:2006kf}. We will note all the 
T-dual quantities with tilde. 

Since the metric is the identity and the $B$ field is only along the base, these two quantities don't change and we have :
\begin{align}
 d\tilde{s}^2=&\mathbb{I}_6\, , & \tilde{B}=&-2|\mu|(-1+\alpha)\tilde{e}^{4}\wedge \tilde{e}^5.
\end{align}
Next we will do the T-duality on the pure spinors. We will first define 
\begin{align}
 \Phi_{+B}=&e^B\Phi_+ \, ,& \Phi_{-B}=&e^B\Phi_- \, . 
\end{align}
Then, the T-duality rules are 
\begin{align}
\label{NSTdual}
 \tilde{\Phi}_{+B}=&-(\iota_1 \Phi_{-B}-e^1\wedge\Phi_{-B})\, ,\\
\tilde{\Phi}_{-B}=&(\iota_1 \Phi_{+B}-e^1 \wedge \Phi_{+B})\, ,
\end{align}
with  $e$ replaced by $\tilde{e}$ in the above expressions. Similarly, the T-dual of the RR fluxes is 
\begin{align}
\label{RRTdual}
 \tilde{F}=e^{-2\tilde{B}}\wedge(\iota_1 -e^1 \wedge )(e^{2B}\wedge F)\, ,
\end{align}
with, again, $e$ replaced by $\tilde{e}$. 

According to the T-duality rules, the component of the $H$-flux with one leg along the duality direction is exchanged with 
some structure constants 
\begin{align}
 f^1_{ab}\leftrightarrow H_{fl~1ab} \, , 
\end{align}
giving the new algebra : $(2|\mu|{35}+2|\mu|{46},0,{25},-{26},-{23},{24})$ and a new $\tilde{H}_{fl}$ :
\begin{align}
 \tilde{H}_{fl}=&2|\mu|(\alpha \tilde{e}^{2}\wedge\tilde{e}^3\wedge\tilde{e}^4-(2-\alpha)\tilde{e}^{2}\wedge\tilde{e}^5\wedge\tilde{e}^6) \, .
\end{align}

We do the transformation $\tilde{e}^1\rightarrow 2|\mu| \tilde{e}^1$ in order to find the solvalgebra we considered
in Section \ref{newsolution}  (with $\epsilon=1$) : $({35}+{46},0,{25},-{26},-{23},{24})$. We can now give explicitly the T-dual solution.
From the transformation of the pure spinors, \eqref{NSTdual},   we can read out the new $\SU(2)$ structure 
\begin{align}
 z=&2|\mu|\tilde{e}^1+i \tilde{e}^2 \, ,\nn \\
 j=&\tilde{e}^{3}\wedge\tilde{e}^6-\tilde{e}^{4}\wedge\tilde{e}^5 \, , \nn \\
\omega_R=&-\tilde{e}^{3}\wedge\tilde{e}^4+\tilde{e}^{5}\wedge\tilde{e}^6 \, , \nn \\
\omega_I=&\tilde{e}^{3}\wedge\tilde{e}^5+\tilde{e}^{4}\wedge\tilde{e}^6 \, , 
\end{align}
and $H$ flux
\beq
H=2|\mu|( \tilde{e}^{2}\wedge\tilde{e}^3\wedge\tilde{e}^4-\tilde{e}^{2}\wedge\tilde{e}^5\wedge\tilde{e}^6)\, . 
\eeq
Similarly, from \eqref{RRTdual} we have
\begin{align}
F_1=&10|\mu|^2 \tilde{e}^1 \, , \nn \\
F_3=& 3|\mu|(\tilde{e}^2\wedge\tilde{e}^4\wedge\tilde{e}^5-\tilde{e}^2\wedge\tilde{e}^3\wedge\tilde{e}^6) \, , \nn \\
F_5=&-6|\mu|^2\tilde{e}^1\wedge\tilde{e}^3\wedge\tilde{e}^4\wedge\tilde{e}^5\wedge\tilde{e}^6 \, . 
\end{align}

This is the solution we had in Section \ref{newsolution} with $j_1=1$, $\epsilon_1=1$, $\omega_1=1$.

\section{Name of the new solution}\label{solvname}

We looked at the classification of six dimensional manifold in \cite{2009arXiv0903.2926B}. One can easily see that our manifold $(35+ \epsilon 46,0,25,-\epsilon 26,-\epsilon23,24)$ has nilradical $\mathfrak{g}_{5.4}$. We rewrite our solution in Bock's notation in order to make the comparisons easier :
\begin{align}
[X_2,X_4]=&X_1 & [X_3,X_5]=&X_1\nonumber\\
[X_2,X_6]=&X_4 & [X_3,X_6]=&-\epsilon X_5\\
[X_4,X_6]=&-\epsilon X_2 & [X_5,X_6]=&X_3\nonumber
\end{align}
We consider manifolds that one can put under the following form :
\begin{align}
[X_2,X_4]=&X_1 & [X_3,X_5]=&X_1\nonumber\\
[X_2,X_6]=&\epsilon_2X_4 & [X_3,X_6]=&\epsilon_3 X_5\\
[X_4,X_6]=&\epsilon_4 X_2 & [X_5,X_6]=&\epsilon_5X_3\nonumber
\end{align}
with $\epsilon_i=\pm 1$. One can see that by redefining $X_6$ in $-X_6$, one can change all the signs of the $\epsilon_i$. So without loss of generality, one can assume that $\epsilon_2=1$. These manifolds are all unimodular (ie $\forall X$, Tr(Ad$_X$)$=0$) solvmanifolds of dimension six with nilradical $\mathfrak{g}_{5.4}$. One can see that up to real redefinitions of the $X$'s, the different cases are :
\begin{center}
\begin{tabular}{|c|c|}\hline
$(\epsilon_3,\epsilon_4,\epsilon_5)$ & $\mathrm{Name}$\\ \hline\hline
$(1,1,1)$ & $\mathfrak{g}_{6.88}$\\ \hline
$(-1,-1,1)$ & $\times$\\ \hline
$(-1,1,-1)$ & $\times$\\ \hline
$(1,-1,-1)$ & $\mathfrak{g}_{6.92}^*$\\ \hline
$(1,1,-1)$ & $\mathfrak{g}_{6.89}^{0,-1,1}=\mathfrak{g}_{6.91}$\\ \hline
$(-1,1,1)$ & $\mathfrak{g}_{6.89}^{0,1,1}=\mathfrak{g}_{6.90}^{0,-1}$\\ \hline

\end{tabular}
\end{center}
The first row of the table corresponds to our solution with $\epsilon=-1$ whereas the second row corresponds to $\epsilon=1$. As one can see, we weren't able to find it in the classification and called it solv$_1$ in the main text. Note that if one permits complex redefinitions of the $X$'s, the first four rows are equivalent and the last two are also equivalent.

\bibliographystyle{utphysmodb}
\bibliography{SUSYAdS}
\end{document}